\documentclass[prb,showpacs,preprintnumbers,amsmath,amssymb]{revtex4}

\usepackage{graphicx}
\usepackage{dcolumn}
\usepackage{bm}

\begin{document}

\title{Magnetic Interference Patterns and Vortices in Diffusive SNS junctions}

\author{J.C. Cuevas and F.S. Bergeret}

\affiliation{Departamento de F\'{\i}sica Te\'orica de la Materia
Condensada, Universidad Aut\'onoma de Madrid, 28049-Madrid, Spain}

\date{\today}

\begin{abstract}
 We study theoretically the electronic and transport properties of a diffusive
 superconductor-normal metal-superconductor (SNS) junction in the presence of a 
 perpendicular magnetic field. We show that the field dependence of the critical 
 current crosses over from the well-known Fraunhofer pattern in wide junctions 
 to a monotonous decay when the width of the normal wire is smaller than the 
 magnetic length $\xi_{\rm H} = \sqrt{\Phi_0/H}$, where $H$ is the magnetic field 
 and $\Phi_0$ the flux quantum. We demonstrate that this behavior is a direct 
 consequence of the magnetic vortex structure appearing in the normal region and 
 predict how such structure is manifested in the local density of states.
\end{abstract}


\maketitle

\emph{Introduction.--}
The study of the modification of the properties of a normal metal in contact 
to superconductors, known as \emph{proximity effect}, has a long 
history~\cite{Deutscher1969}. In the last years there has been
a renewed interest in this subject because new experimental techniques 
have allowed resolving properties on smaller length scales and
very low temperatures~\cite{Pannetier2000}. Although many electronic and 
transport properties of hybrid SN structures are now well understood, the
situation is less satisfactory when dealing with the magnetic field dependence
of those properties. A few years ago, Heida \emph{et al.}~\cite{Heida1998} measured 
the critical current as a funcion of a perpendicular magnetic field in ballistic 
SNS junctions of comparable length and width and found a periodicity close to $2\Phi_0$,
where $\Phi_0=h/2e$ is the flux quantum, instead of the standard $\Phi_0$ of 
the Fraunhofer pattern~\cite{Barone1982}. This was qualitatively explained in 
Refs.~[\onlinecite{Barzykin1999,Ledermann1999}] in terms of the classical 
trajectories associated with current-carrying Andreev states in a normal clean wire. 
In the case of diffusive junctions, numerous experiments have shown that in wide 
junctions the critical current exhibits a Fraunhofer-like pattern~\cite{Clarke1969,
Nagata1982}. However, very recent experiments in junctions where the width is 
comparable to the superconducting coherence length have shown a monotonous decay 
of the critical current with field, i.e. the absence of magnetic interference 
patterns~\cite{orsay2007}. The unified description of these two very different
behaviors is a basic open problem.

In this Letter we show that the solution to the previous puzzle is closely
related to the issue of the formation of a magnetic vortex structure in the
normal conductor. Vortex matter in mesoscopic superconductors has been also 
a very active subfield in superconductivity in the last years~\cite{Moshchalkov1999}. 
It has been shown that basic properties such as critical fields~\cite{Moshchalkov1995} 
and the magnetization~\cite{Geim1997} depend crucially on the size and topology 
of the mesoscopic samples, which in turn determine the vortex structure. There is 
also a great interest in the study of nucleation of superconductivity and vortex 
matter in hybrid structures~\cite{Moshchalkov2006}. However, little attention has 
been paid to the formation of vortices inside non-superconducting materials. Our 
goal here is to answer the following fundamental questions: Is it possible to induce a 
vortex structure in a normal wire by proximity to a superconductor?, if so, what are 
the properties of such \emph{proximity vortices} and their influence on the Josephson
effect? For this purpose, we have studied a diffusive SNS junction in the 
presence of a perpendicular magnetic field. By solving the two-dimensional Usadel 
equations~\cite{Usadel1970}, we are able to describe the electronic properties for 
arbitrary length, $L$, and width, $W$, of the normal wire. We find that a magnetic 
vortex structure may develope in the normal metal. These vortices have similar 
properties to those in the mixed state of a type II superconductor~\cite{Abrikosov}. 
The consequence of this vortex structure is the
appearance of an interference pattern in the critical current that tends to the
Fraunhofer pattern in the wide-junction limit ($W\gg \xi_{\rm H}=\sqrt{\Phi_0/H}$), 
and also a modulation of the local density of states in the normal wire. On the 
contrary, when $W$ is comparable or smaller than $\xi_{\rm H}$, the formation 
of vortices is not favorable and the field acts as a pair-breaking mechanism 
which suppresses monotonously the critical current. Our results not only solve the 
puzzle described above, but also illustrate the richness of the vortex physics in 
hybrid structures.

\emph{Quasiclassical formalism.--}
We consider a SNS junction, where N is a diffusive normal metal of length $L$
and width $W$ coupled to two identical superconducting reservoirs with gap $\Delta$.
The junction is subjected to an uniform external field ${\bf H} = H \hat z$ perpendicular
to the normal film lying in the $xy$-plane, where $x \in [0,L]$ and $y \in [-W/2,
W/2]$. For the sake of simplicity, we assume that the thickness of the normal wire 
is smaller than the London penetration depth, which means that the field penetrates 
completely in the normal region. In order to describe the electronic properties of 
these junctions we use the quasiclassical theory of superconductivity in the diffusive
limit~\cite{Usadel1970,Larkin1986}, where the mean free path is much smaller
than the  coherence length, $\xi=\sqrt{\hbar D/ \Delta}$, $D$ being the diffusion 
constant of the normal metal. In equilibrium situations like the
one considered here, this theory can be formulated in terms of momentum averaged 
retarded Green functions $\hat G^R({\bf R},\epsilon)$, which depend on 
position ${\bf R}$ and energy $\epsilon$. This propagator is a $2\times 2$ matrix 
in electron-hole space
\begin{equation}
\label{retarded}
\hat G^{R} = \left( \begin{array}{cc}
{\cal G}^{R} & {\cal F}^{R} \\
\tilde {\cal F}^{R}  & \tilde {\cal G}^{R}
   \end{array} \right) ,
\end{equation}
which satisfies the stationary Usadel equation, which in the N region
reads~\cite{Larkin1986}
\begin{equation}
\label{usadel-eq}
\frac{\hbar D}{\pi} {\bf \nabla} \left( \hat G^R {\bf \check \nabla}
\hat G^R \right) + \; \epsilon [ \hat \tau_3, \hat G^R ] =
\frac{i e D}{\pi} {\bf A} [ \hat \tau_3, \hat G^R 
{\bf \check \nabla} \hat G^R ]  .
\end{equation}
\noindent
Here, ${\bf A}$ is the vector potential, ${\bf \check \nabla} = {\bf \nabla} 
\hat 1 - (ie/\hbar){\bf A} \hat \tau_3$, $\hat{\tau}_3$ is the Pauli matrix 
and the Coulomb gauge (${\bf \nabla} {\bf A} = 0$) has been already used. 
 Eq.~(\ref{usadel-eq}) is supplemented by the 
normalization condition $(\hat G^R)^2 = -\pi^2 \hat 1$ and proper boundary 
conditions. For the SN interfaces we use the boundary conditions introduced 
in Ref.~[\onlinecite{Nazarov1999}], which allow us to describe the system for 
arbitrary transparency. For the metal-vacuum borders of the normal wire we 
impose that the current density in the $y$-direction vanishes at 
$y=\pm W/2$~\cite{note1}. In general, the Usadel equation has to be solved 
together with the Maxwell equation ${\bf \nabla} \times {\bf H} = \mu_0 
{\bf j}$ in a self-consistent manner. However, we are interested here in the 
case where the width $W$ is smaller than the Josephson penetration length 
$\lambda_{\rm J}=\sqrt{\hbar /2 \mu_0 e j_c d}$, where $j_c$ is the critical 
current per unit area and $d$ is the effective length of the junction including 
the London penetration depths in the leads. In this case one can ignore the 
screening of the magnetic field by the Josephson currents and the field is 
equal to the external one~\cite{Barone1982}.

The physical properties we are interested in can be conveniently expressed
in terms of the Usadel-Green functions. Thus for instance, the local density of 
states is given by $\rho({\bf R},\epsilon) = - \mbox{Im} {\cal G}^R({\bf R},
\epsilon)/\pi$.  To quantify the superconducting correlations we use
the pair correlation function defined as $F({\bf R}) = (1/4\pi i) \int d\epsilon 
({\cal F}^{R} -{\cal F}^{A}) \tanh (\beta \epsilon/2)$, where $\beta = 1/k_{\rm B}T$.
This function, apart from the atractive coupling constant, is the pair potential 
in a superconductor and it is non-zero inside the normal metal due to the 
proximity effect. Finally, the supercurrent density in the junction can be 
written as
\begin{equation}
{\bf j}({\bf R}) = \frac{\sigma_{\rm N}}{4\pi^2 e} \int^{\infty}_{-\infty} d\epsilon
\; \tanh \left( \frac{\beta \epsilon}{2} \right) 
\mbox{Re} \left\{ {\cal F}^{R} {\bf \nabla} \tilde {\cal F}^{R} -
\tilde {\cal F}^{R} {\bf \nabla} {\cal F}^{R} +
\frac{4ie}{\hbar} {\bf A} {\cal F}^{R} \tilde {\cal F}^{R} \right\} ,
\end{equation}
where $\sigma_{\rm N}$ is the normal state conductivity. The net current is
obtained integrating $j_x$ across the $y$-direction.

Eq.~(\ref{usadel-eq}) constitutes a set of coupled second-order nonlinear 
partial differential equations, whose resolution is a formidable task. In 
general, one has to resort to numerical methods. However, one can get 
analytical insight in several limiting cases. By choosing the gauge ${\bf A} 
= -Hy \hat x$, one can identify in Eq.~(\ref{usadel-eq}) the length 
$\xi_{\rm H} = \sqrt{\Phi_0/H}$ as the characteristic variation scale of the 
Green functions in the transversal direction due to the magnetic field. We consider first the case where 
the wire width $W$ is smaller than $\xi_{\rm H}$. In this case 
the Green functions do not vary considerably in the $y$-direction and one can 
average Eq.~(\ref{usadel-eq}) over this direction leading to the   
one-dimensional equation
\begin{equation}
\label{1D}
\frac{\hbar D}{\pi} \partial_x \left( \hat G^R \partial_x
\hat G^R \right) + \epsilon [\hat \tau_3, \hat G^R]
= \frac{\Gamma_{\rm H}}{2 \pi} [ \hat \tau_3 \hat G^R
\hat \tau_3 , \hat G^R ] ,
\end{equation}
where $\Gamma_{\rm H} = D e^2 H^2 W^2/(6 \hbar)$ is a depairing energy,
which in terms of the Thouless energy, $\epsilon_{\rm T} = \hbar D /L^2$, can
be written as $\Gamma_{\rm H} = \epsilon_{\rm T} (\pi \Phi / \sqrt{6} \Phi_0)^2$, 
where $\Phi = HLW$ is the flux enclosed in the junction. These equations describe 
the effect of a pair-breaking mechanism, such as magnetic impurities, that has
been studied extensively in Ref.~[\onlinecite{Hammer2007}]. The other analytic 
case is the limit of a wide junction where $W \gg L, \xi_{\rm H}$. In this limit
one can neglect the terms containing the derivatives with respect to the 
$y$-coordinate. The field also disappears from the equation and its only effect 
is to change the superconducting phase difference $\phi$ into the gauge-invariante 
combination $\gamma = \phi - 2\pi(\Phi / \Phi_0) y/W$. With this result in mind, 
it is easy to anticipate, in particular, that critical current exhibits a 
Fraunhofer-like pattern in this limit. 

\begin{figure}[t]
\includegraphics[width=\columnwidth]{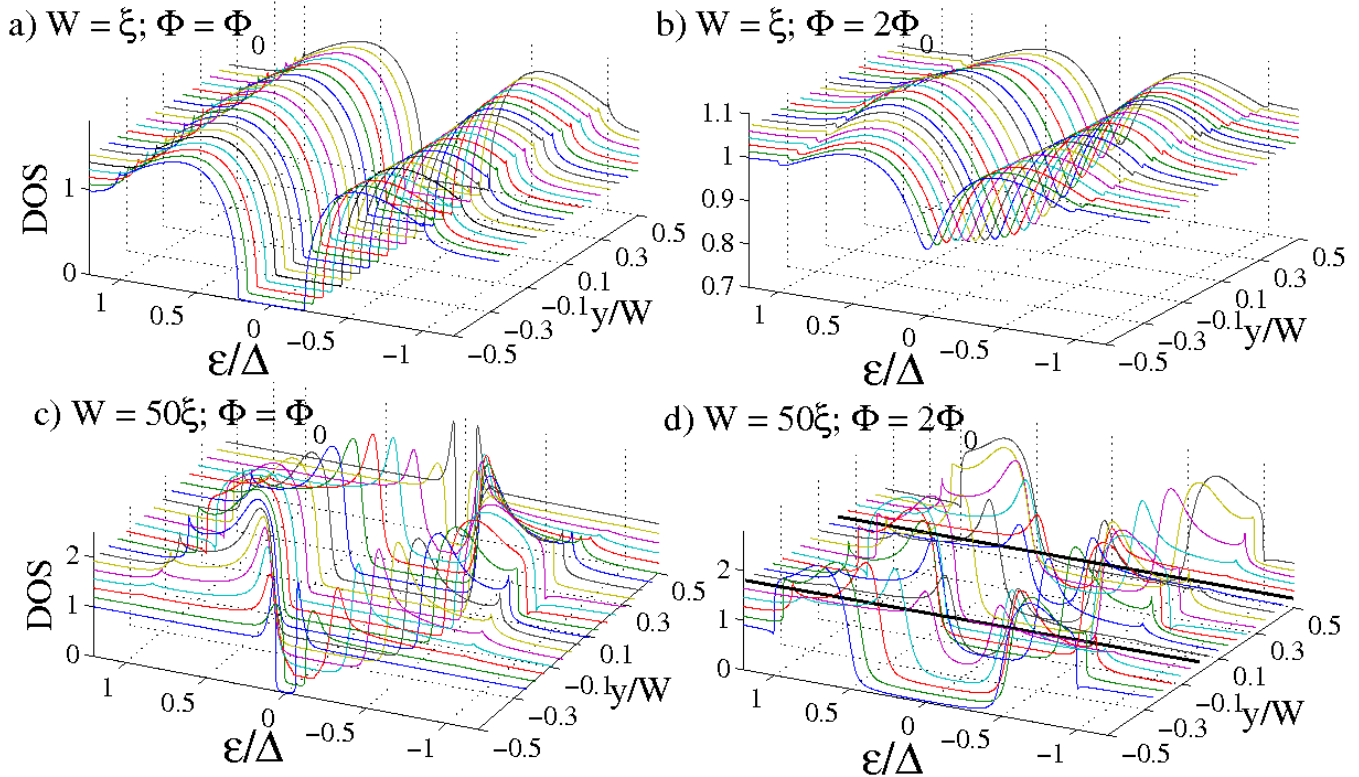}
\caption{\label{dos} (Color online) Local density of states as a function of 
the energy in the middle of a wire of length $L=2\xi$ for two different values
of the magnetic flux. The upper panels correspond to a wire width $W=\xi$ and
the lower ones to $W=50\xi$. The different curves correspond to different values 
of the $y$-coordinate. We have assumed perfect transparency for the interfaces 
and a phase difference $\phi=0$. In panel (d) we have used thicker lines to 
highlight the curves where the DOS is equal to the normal state one.}
\end{figure}

\emph{Discussion of the results.--}
We start by analyzing the local density of states (DOS) in the normal wire. 
In the absence of field the main feature is the presence of a minigap, 
$\Delta_{\rm g}$~\cite{Golubov1988,Belzig1996,Zhou1998}. This minigap is the same 
throughout the normal wire and for perfect transparency scales as $\Delta_{\rm g} 
\sim 3.1 \epsilon_{\rm T}$ in the limit $L \gg \xi$. In Fig.~\ref{dos} we show 
the local DOS in the middle ($x=L/2$) of a wire of length $L=2\xi$ for two 
different values of the width and the magnetic flux. Notice that for $W=\xi$ 
(see upper panels), the local DOS is practically independent of the $y$-coordinate. 
Moreover, when the field is not very high, there is a clear minigap (see upper 
left panel), which closes at higher fields (see upper right panel). As one can 
see in the lower panels, when $W \gg L$, the local DOS is strongly modulated 
along the $y$-direction. For low fields ($\Phi < \Phi_0$), the minigap is still 
open throughout the wire, but for higher fields the minigap changes in a periodic fashion 
from its maximum value (equal to the value in the absence of field) to exactly zero 
at well-defined positions where the DOS is the normal state one. This 
situation is very similar to the mixed state of a type II superconductor, where 
the system becomes normal in the vortex cores.

These results are in agreement  with the limiting cases discussed 
above. If the wire is narrow the magnetic field acts as a pair-breaking
mechanism with depairing energy $\Gamma_{\rm H}$. It is well-known that the minigap 
is reduced by such mechanisms~\cite{Belzig1996,Hammer2007} and, in particular, it 
closes at a critical value $\Gamma^{\rm C}_{\rm H} = \pi^2 \epsilon_{\rm T} /
2$~\cite{Crouzy2005}, i.e. in our case at a critical flux $\Phi^{\rm C} = \sqrt{3} 
\Phi_0$. This explains the results for $W=\xi$. To understand the results 
for $W=50\xi$, we remind that in the wide limit  the magnetic field only enters 
in the gauge-invariant phase difference $\gamma$. 
It has been shown that in the absence of field the minigap decreases monotonously 
as the phase difference increases and it closes when the phase is equal to 
$\pi$~\cite{Zhou1998}. Bearing this in mind, one can easily understand the results 
of Fig.~\ref{dos}(c,d). When $\gamma = 0$ the minigap is completely open reaching the
value in the absence of field. However, when $\gamma = \pi$ the minigap closes.
For $\Phi=2\Phi_0$ and $\phi = 0$, the phase $\gamma$ takes the values $\mp \pi$
at $y/W = \pm 1/4$, which explains why the two thick curves in panel (d)
correspond to normal state DOS.

\begin{figure}[t]
\includegraphics[width=\columnwidth,clip]{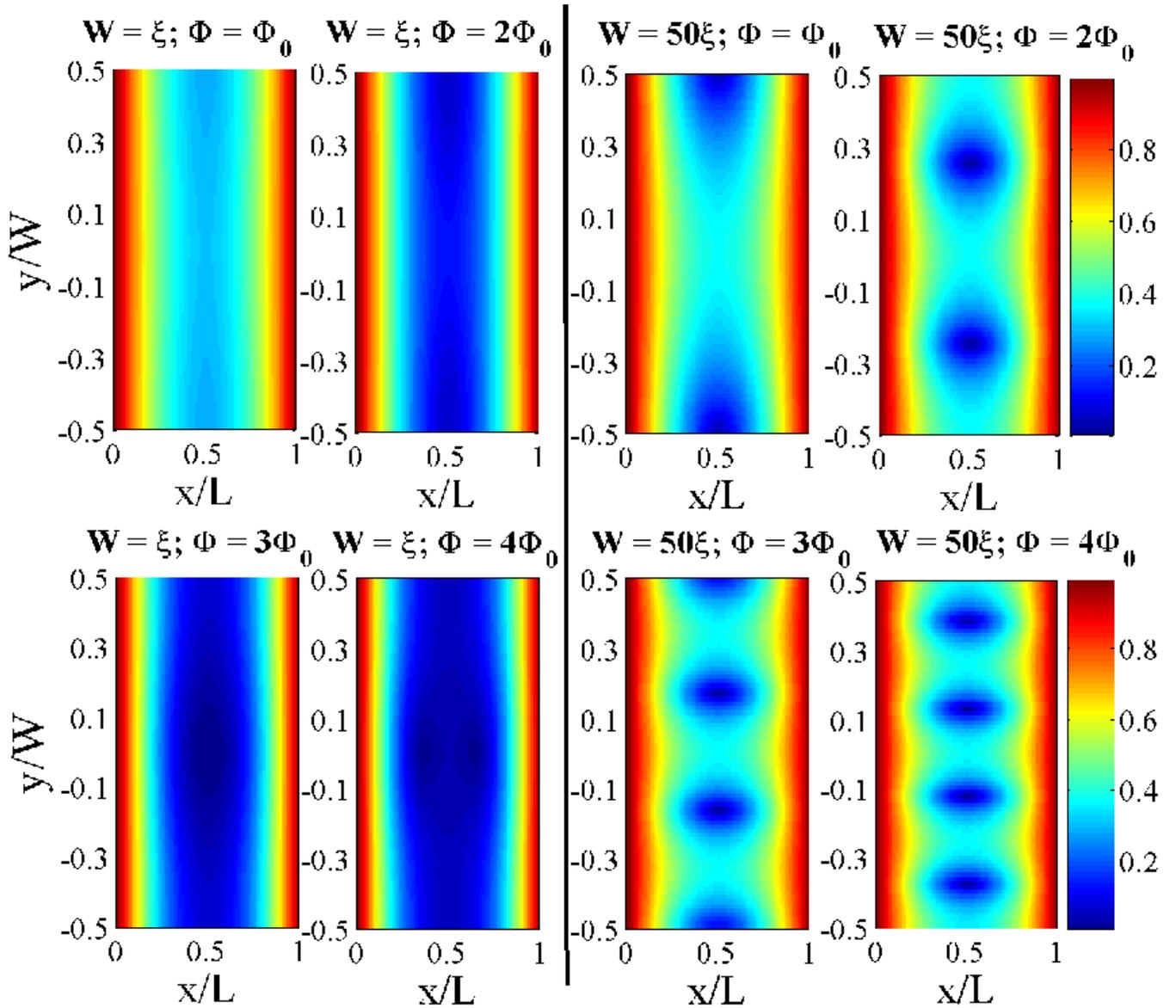}
\caption{\label{vortices} (Color online) Spatial map of the modulus of the 
pair correlations, $|F({\bf R})|$, for $L=2\xi$ and $\phi=0$. The
different panels correspond to different values of the width $W$ and the
magnetic flux $\Phi$, as indicated in the graphs. $|F({\bf R})|$ has been 
normalized to its value inside the electrodes, the temperature is 
$k_{\rm B} T = 0.01\Delta$ and perfect transparency was assumed.} 
\end{figure}

The peculiar DOS suggests the presence of vortices in 
the normal wire. To confirm this idea, we have analyzed the pair correlation 
function, $F({\bf R})$. In Fig.~\ref{vortices} we show  a color-coded map of
the modulus of this function throughout the normal wire for the same values 
of $L$ and $W$ as in Fig.~\ref{dos}. All the panels show that $F$ diminishes 
towards the center of the wire, which simply reflects the decay of the 
superconducting correlations inside the normal wire. The main difference is 
the modulation along the $y$-direction. In the case $W=\xi$, at low fields 
(see panel for $\Phi = \Phi_0$) $F$ is still finite everywhere, while for higher 
fields it can be very small in the center of the wire, but with practically 
no modulation. The situation changes drastically for $W=50\xi$, where one can 
clearly see the appearance of a linear array of vortices located on $x=L/2$ with
normal cores where $F$ vanishes. The number of vortices depends simply on the 
number of flux quanta in the junction. It is important to distinguish these 
vortices from the so-called Josephson vortices that take place in much wider 
junctions where $W>\lambda_J$ and  the self-field effects play a crucial role~\cite{Barone1982}. 
An important difference is that the Josephson vortices do not have normal cores and their size is comparable to $\lambda_J$.

It is possible to get an analytical insight into the vortex structure by
using the linearized Usadel equations in the wide-junction limit. In this case 
one can show that the position of the zeros of the pair correlation function are
given by 
\begin{equation}
\label{zeroes}
x=\frac{L}{2} \;\; \mbox{and} \;\; \phi - 2\pi \left( \frac{\Phi}{\Phi_0} \right) 
\frac{y}{W} = (2m+1) \pi ,
\end{equation}
where $m=0,\pm 1,...$ and $y \in [-W/2,W/2]$. This means that the vortex 
cores are located exactly on the middle of the wire forming a regular linear 
array along the $y$-direction and they are separated by a distance $\Phi_0/HL$. 
Thus, for the case $W=50\xi$ in Fig.~\ref{vortices} this condition tells us
that for $\Phi = 4\Phi_0$ there are four vortex cores located on $y/W = \pm 1/8,
\pm 3/8$, which are the positions that one can read off from Fig.~\ref{vortices}.
Notice also that according to Eq.~(\ref{zeroes}) the phase $\phi$ simply shifts 
rigidly the line of vortices along the $y$-direction. Thus, measurements of the 
local DOS at the outer interfaces ($y=\pm W/2$) changing the supercurrent 
through the junction should show an oscillatory behaviour. On the other
hand, the normal cores are surrounded by circulating currents (not shown here), which
vanish exactly at the cores. Moreover, from the analytical solution of the linearized Usadel
equation in the wide-junction limit and from the numerical results for arbitrary cases, one can easily show that the phase of the
pair correlation changes in $2\pi$ around the cores, {\it i.e} each vortex has a 
unit topological charge.  In short, the main difference with the
usual vortices in a bulk superconductor of type II is that they
are arranged in one-dimensional array instead of forming a two-dimensional
lattice and due to the confining geometry they do not possess a rotational 
symmetry~\cite{Abrikosov}.

\begin{figure}[t]
\begin{center}
\includegraphics[width=\columnwidth,clip]{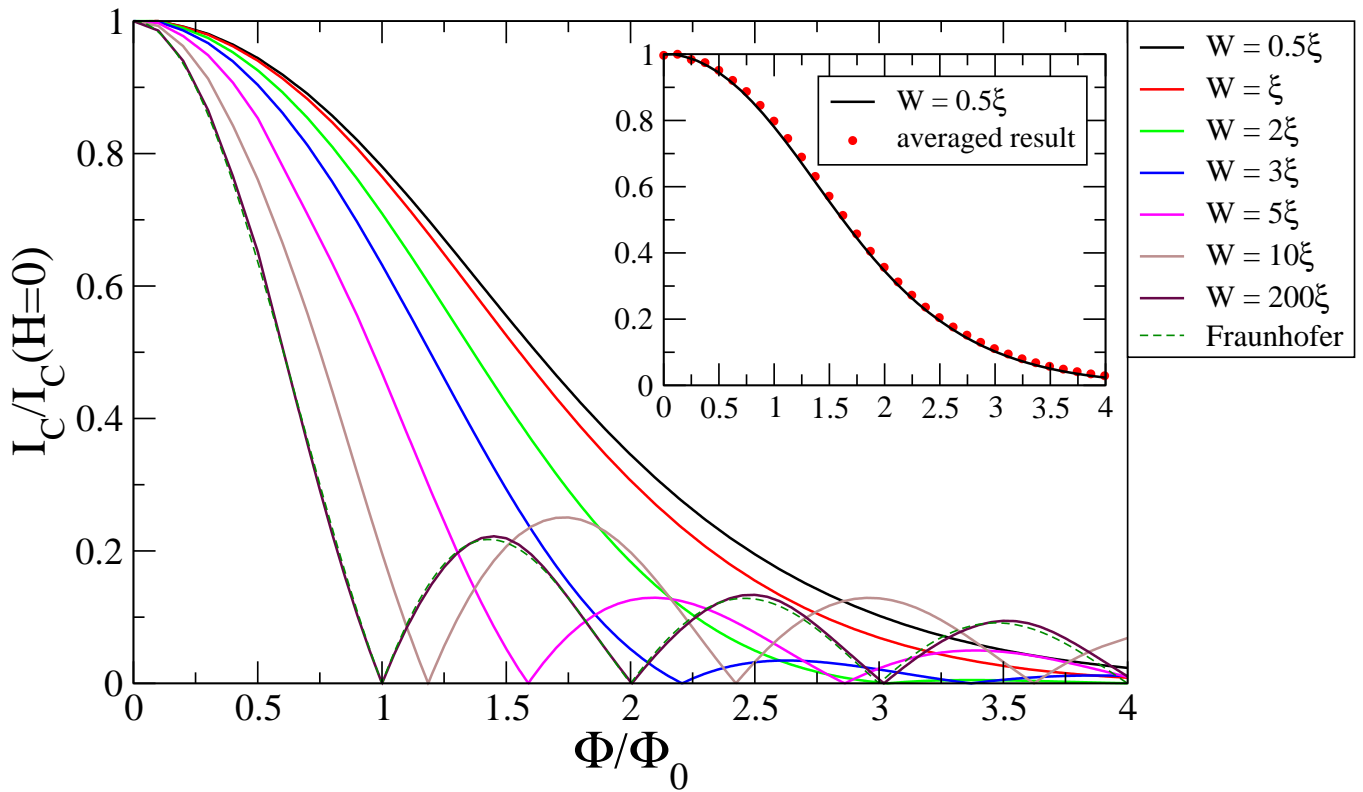}
\caption{\label{Fraunhofer}(Color online) Critical current normalized by the zero-field
value vs. magnetic flux for a wire length $L=8\xi$, perfect transparent 
interfaces and $k_{\rm B}T = 0.01\Delta$. The different curves correspond 
to different values of the width of the wire, $W$. The dashed line 
shows the standard Fraunhofer pattern given by $\sin(\pi \Phi / \Phi_0)/
(\pi \Phi / \Phi_0)$. The inset shows for $W=0.5\xi$ the comparison between 
the exact result and the approximation of Eq.~(\ref{1D}) for the 
narrow-junction limit.}
\end{center}
\end{figure} 

We discuss finally the magnetic field dependence of the critical current. 
In Fig.~\ref{Fraunhofer} we show an example for $L=8\xi$, 
which is a typical value in the experiments~\cite{orsay2007}, and different 
values of $W$. Notice that for small values of $W$, the critical 
current decays monotonously. This is simply due to the fact that in this limit
no vortices appear and the field suppresses progressively 
the superconductivity in the normal wire~\cite{Hammer2007}. 
Indeed, as we show in the inset of Fig.~\ref{Fraunhofer}, in this limit 
Eq.~(\ref{1D}) describes quantitatively the field dependence. As
the width increases the vortex structure discussed above appears and as a consequence one observes the appearance of an interference pattern
where the critical current vanishes at certain values of the magnetic 
flux.  Notice that these values are clearly larger than 
$\Phi_0$ for intermediate widths and the patterns are not ``periodic". 
Only in the limit $W \gg \xi_{\rm H}, L$ one obtains a regular pattern with zeros at 
multiples of $\Phi_0$, recovering the Fraunhofer pattern~\cite{Barone1982}.
These results 
explained in an unified manner the different behaviours observed 
experimentally~\cite{Clarke1969,Nagata1982,orsay2007}, which at first 
glance seemed to be contradictory.

Finally, we have studied systematically the role of the length $L$ in 
the crossover from the narrow-junction to the wide-junction behavior. We have 
found that as $L$ increases this transition occurs at larger values of $W$.
This confirms the fact that the condition for the appearence of an interference 
pattern, i.e. zeros in the critical current, is given, roughly speaking, by 
$W>\xi_{\rm H}$, which is equivalent to $W/L>\Phi_0/\Phi$. The standard 
Fraunhofer pattern is approach when $W \gtrsim L$.

\emph{Conclusions.--}
We have studied a diffusive SNS junction in the presence of a perpendicular 
magnetic field. We have shown that the appearance of magnetic interference patterns 
in the critical current is intimately linked to the formation of a vortex array in 
the normal wire. Our results provide an unified description of the critical current
for arbiratry width of the junctions and solve the puzzle put forward
by recent experiments~\cite{orsay2007}. Our work also paves the way to study 
the vortex matter in a great variety of hybrid structures like the recently 
introduced superconducting graphene junctions, where a standard Fraunhofer
pattern has been observed~\cite{Heersche2007}.

We would like to thank Sophie Gu\'eron and H\'el\`ene Bouchiat for numerous discussions
about their experiments which motivated this work. It is also a pleasure to acknowledge 
useful discussions with Philippe Joyez, Hugues Pothier, T.M. Klapwijk, S. Russo, 
A.F. Morpurgo and J.G. Rodrigo. This work has been financed by the Spanish CYCIT 
(contract FIS2005-06255). F.S.B. acknowledges funding by the Ram\'on y Cajal program.


\end{document}